\documentclass[preprint]{aastex}
\usepackage{epsf}
\usepackage{emulateapj5}
\usepackage{onecolfloat}
\usepackage{amsmath}

\def\gtsim {\lower .1ex\hbox{\rlap{\raise .6ex\hbox{\hskip .3ex
        {\ifmmode{\scriptscriptstyle >}\else
                {$\scriptscriptstyle >$}\fi}}}
        \kern -.4ex{\ifmmode{\scriptscriptstyle \sim}\else
                {$\scriptscriptstyle\sim$}\fi}}}
\newcommand{\beq}{\begin{equation}}
\newcommand{\eeq}{\end{equation}}

\def\etal{{et al.~}}
\def\Mo{{\rm M_\odot}}
\def\Mpc{\ {\rm Mpc}}
\def\kpc{\ {\rm kpc}}
\def\pc{\ {\rm pc}}
\def\kms{{\ }{\rm km}\,{\rm s}^{-1}}
\def\LCDM{$\Lambda$CDM}

\begin{document}
\submitted{The Astrophysical Journal, accepted}
\vspace{1mm}
\slugcomment{{\em The Astrophysical Journal, accepted}}

\twocolumn[
\lefthead{DENSITY PROFILES OF COLD DARK MATTER SUBSTRUCTURE}
\righthead{KAZANTZIDIS ET AL.}

\title{DENSITY PROFILES OF COLD DARK MATTER SUBSTRUCTURE: \\
IMPLICATIONS FOR THE MISSING-SATELLITES PROBLEM}

\author{Stelios Kazantzidis\altaffilmark{1}, Lucio Mayer, Chiara Mastropietro, J\"urg Diemand,
Joachim Stadel, and Ben Moore}
\vspace{1mm}
\affil{Institute for Theoretical Physics, University of Z\"urich, CH-8057 
Z\"urich, Switzerland}

\begin{abstract}
 The structural evolution of substructure in cold dark matter (CDM) 
 models is investigated combining ``low-resolution'' satellites from cosmological $N$-body 
 simulations of parent halos with $N=10^7$ particles with high-resolution 
 individual subhalos orbiting within a static host potential.
 We show that, as a result of mass loss, convergence 
 in the central density profiles requires the initial satellites to be resolved 
 with $N=10^7$ particles and parsec-scale force resolution. 
 We find that the density profiles of substructure halos can be well fitted with
 a power-law central slope that is unmodified by tidal forces even after the tidal
 stripping of over 99\% of the initial mass and an exponential cutoff in the outer parts.
 The solution to the missing-satellites problem advocated by Stoehr \etal in 2002
 relied on the flattening of the dark matter halo central density cusps by gravitational 
 tides, enabling the observed satellites to be embedded within dark halos with maximum 
 circular velocities as large as $60\kms$. In contrast, our results suggest that tidal 
 interactions do not provide the mechanism for associating the dwarf spheroidal 
 satellites (dSphs) of the Milky Way with the most massive substructure halos expected in a 
 CDM universe. Motivated by the structure of our stripped satellites, we
 compare the predicted velocity dispersion profiles 
 of Fornax and Draco to observations, assuming that they are 
 embedded in CDM halos. We demonstrate that models with isotropic and
 tangentially anisotropic velocity distributions for the stellar component 
 fit the data only if the surrounding dark matter halos
 have maximum circular velocities in the range $20-35 \kms$.
 If the dSphs are embedded within halos this large then the overabundance
 of satellites within the concordance {\LCDM} cosmological model is
 significantly alleviated, but this still does not provide the entire solution. 
\end{abstract}

]

\altaffiltext{1}{E-mail: stelios@physik.unizh.ch}

\section{INTRODUCTION}
\label{section:introduction}
The concordance Lambda cold dark matter 
({\LCDM}) cosmological model for structure formation has been remarkably 
successful in explaining most of the properties of the 
universe on large scales and at various cosmic epochs.
Recent results from microwave background experiments
and large redshift surveys have highlighted the ability of this model 
to reproduce observations as diverse as the abundance and clustering of galaxies 
and clusters and the statistical properties of the Ly$\alpha$ forest 
within constraints set by measurements of the primordial fluctuation spectrum,
observations of distant Type Ia supernovae, and gravitational lensing statistics
\citep[e.g.,][]{phillips_etal01,jaffe_etal01,percival_etal01,hamilton_tegmark02,
croft_etal02,bahcall_etal03,tonry_etal03,spergel_etal03}.
However, on non-linear scales the {\LCDM} model has been neither 
convincingly verified nor disproved, and several outstanding issues 
remain unresolved. 

Specifically, the rotation curves of dwarf and low surface 
brightness galaxies are better fitted by shallower
dark matter density profiles and lower
concentrations than those predicted by the standard model
\citep[e.g.,][]{flores_primack94,moore94,burkert95,mcgaugh_deblok98,deblok_etal01,
deblok_mcgaugh_rubin01,deblok_bosma02,mcgaugh_etal03,simon_etal03}.
However, cuspy dark matter distributions may be consistent with the rotation
curve data once systematic uncertainties are considered, such as 
non-circular motions due to the presence of a bar 
\citep{valenzuela_klypin03,rhee_etal03}. Fast rotating bars require 
dark matter densities on galactic scales significantly lower than 
the theoretical predictions \citep{debattista_sellwood00}
and fluid dynamical models based on observations of barred galaxies support the same conclusion \citep{weiner_etal01}.
The inner Galaxy mass profile is marginally consistent with having 
dark matter in the central region \citep{binney_evans01}.
On cluster scales, gravitational lensing observations are similarly 
suggestive of a discordance between the measured shallow
dark matter density inner slopes and the predicted by
numerical simulations cusps \citep{sand_etal02,sand_etal04}.
Nevertheless, hydrodynamical calculations of cluster formation have yet to 
be carried out and compared in detail with the observational data.

Among the most puzzling discrepancies on small scales is the 
so-called substructure problem. {\LCDM} 
predicts a number of subhalos within the Local Group with maximum circular 
velocities in the range $V_{\rm max} \sim 10-30 \kms$, which is
about $2$ orders of magnitude higher than the total number of 
observed satellite galaxies \citep{kauffmann_etal93, moore_etal99,klypin_etal99}.
A number of possible solutions has been proposed to alleviate this problem. 
One class of solutions is related to radical modifications of the 
fundamental {\LCDM} paradigm itself,
including allowing for a finite dark matter particle
self-interaction cross section that enhances the satellite destruction 
within galactic halos \citep{spergel_steinhardt00}, reducing the small-scale power 
with a warm dark matter candidate 
\citep[e.g.,][]{avila_reese_etal01,eke_etal01,bode_etal01},
changing the shape of the primordial power spectrum
\citep{kamionkowski_ liddle00,zentner_bullock02}, 
introducing an interaction between dark matter particles and photons 
\citep{boehm_etal02} or resorting to the decay of a charged particle
to suppress the small-scale power spectrum \citep{sigurdson_kamionkowski03}.

A second class of solutions relies on the inability of low mass
satellites to form stars, by either supernova feedback, photoionization, or 
reionization. Indeed, baryon dissipation, star formation, and radiative feedback mechanisms
must have a decisive effect on the properties of the final observed system, 
and it has been suggested that suppression of gas collapse and/or cooling by a 
photoionizing background at high redshift
can dramatically reduce the number of visible satellites with $V_{\rm vir} < 50\kms$, 
possibly reconciling the observations with the model predictions
\citep{bullock_etal00,benson_etal02a,benson_etal02b}.
Nonetheless, these semi-analytical calculations 
adopt a simple description of the radiation physics 
and disagree with recent numerical simulations showing that even 
very small dwarf galaxies with circular velocities of the order 
of $V_{\rm vir} \sim 30\kms$ or even lower can self-shield themselves 
from the UV background and form some stars \citep{susa_umemura03}.
Feedback processes may also violate the strong correlation between baryonic 
mass and virial velocity, as emphasized by \citet{mayer_moore03}.

Recently, another way of relieving the missing satellites problem
has been advocated by \citet[hereafter S02]{stoehr_etal02}.
These authors argue that the observed Galactic satellites are hosted by
the most massive substructures within a given CDM halo.
In this case, feedback processes suppress star formation in all smaller objects, 
and the ``luminosity function'' of the dwarf spheroidal galaxies (dSphs)
is consistent with the ``mass function'' of the subhalos. 
This represents a somewhat simpler scenario than forming
stars in a random 10\% of the satellites over a larger mass range.
However, in order to achieve this renormalization of the observed satellite 
mass/circular velocity function, S02 claimed that the observed velocity dispersions 
of the dSphs do not correlate with the maximum circular velocities, $V_{\rm max}$,
of their dark matter halos in the way originally 
assumed by \citet{moore_etal99} and \citet{klypin_etal99} 
when they identified the substructure problem.

As an illustrative example, consider the case of the dSph Fornax, with an
observed stellar central velocity dispersion of
$\sigma_{\star} \sim 10 \kms$ \citep{mateo98}. The circular velocity
of the surrounding dark matter halo at the location of the dwarf 
can be of the order of $V_{\rm c} = a \sigma_{\star} \sim 15-20 \kms$, 
depending on the particular assumptions made to infer such quantities 
from observations. For example, an isothermal halo model with a 
flat circular velocity profile, as adopted by \citet{moore_etal99}, would
yield $V_{\rm c}=\sqrt{2}\sigma_{\star}=14.1 \kms$. 
Provided that the circular velocity profile is 
still slowly rising in the region where the visible dwarf galaxy resides,
$V_{\rm c}$ may substantially underestimate $V_{\rm max}$.
This would clearly allow the dwarf galaxies 
to be embedded in dark halos with very large maximum circular velocities in 
the range of $V_{\rm max}= 50-60 \kms$.
However, such slowly rising circular velocity profiles seem to be in 
notable disagreement with previous studies suggesting that the standard 
{\LCDM} halos corresponding to the dSphs are expected to be 
very concentrated, which naturally leads to steeply rising circular velocity profiles
\citep{bullock_kravtsov_weinberg01}.

S02 suggested that the satellites experience significant mass 
redistribution in their centers as a result of tidal interactions, 
leading to shallower inner density profiles and smaller effective concentrations
than those of comparable isolated halos. However, the subhalos in their 
cosmological simulations contained just a few thousand particles and the
softening lengths they used were comparable to the entire extent of some of the dSphs
they wanted to resolve. These authors corroborated their results by comparing 
them with the higher resolution $N$-body simulations of 
\citet[hereafter H03]{hayashi_etal03}, who employed 
individual cuspy model satellites disrupting in a static primary potential and found the 
same shallow inner density profiles as a result of tides.
Nonetheless, the later models have the major shortcoming that they do not constitute 
equilibrium configurations, since they are constructed by approximating
the exact velocity distribution at any given point in space with a Maxwellian. 
When evolved in isolation, these models relax rapidly to an inner density slope 
much shallower than the originally intended one, 
and any results obtained should be treated with care 
(Kazantzidis, Magorrian, \& Moore 2004).

In this paper, we investigate how a satellite's internal structure
responds to tidal interactions, revisiting the solution to the missing-satellites 
problem proposed by S02. Our study uses $N$-body cosmological 
simulations of renormalized systems, in which the subhalos have up to $15$ times 
more particles than those used by S02. Note, however,
that \citet{stoehr_etal03} confirmed the results of S02, using a simulation with a
factor of $9$ more particles. Furthermore, we
investigate the tidal disruption of individual cuspy substructure halos, 
employing $N=10^7$ particles, $100$ times the mass resolution of H03, 
to minimize numerical effects \citep{moore_etal96,diemand_etal04a}.
Our primary goal is to examine the change in the internal structure of
the substructure at scales comparable to the actual sizes of the luminous 
dwarf galaxies as they suffer tidal shocks and gravitational stripping.
The models we adopt for the individual satellite 
simulations are self-consistent realizations and thus ideal for the sensitivity of this
particular study. As we illustrate below, tides do not modify 
the central density structure of cuspy satellites and therefore the missing 
satellites problem can not be solved in this way.

This paper is organized as follows. In \S~\ref{section:num_sim}, we discuss
our cosmological and individual satellite simulations and present 
our results regarding the internal structural evolution of the substructure.
In \S~\ref{section:dSph_kinematics}, we model the kinematics 
of the Draco and Fornax dSphs on the basis of the findings of our simulations,
and in \S~\ref{section:discussion} we discuss the implications of our results.
Finally, in \S~\ref{section:summary} we summarize our main findings and 
conclusions.

\section{NUMERICAL SIMULATIONS}
\label{section:num_sim}

All the simulations carried out in this paper were performed
with PKDGRAV \citep{stadel01}, a multi-stepping, parallel,
$N$-body tree code that
uses a spline kernel softening such that the force is completely Keplerian at twice
the quoted softening lengths.  We used an adaptive,
kick-drift-kick (KDK) leapfrog integrator, and the individual particle
time-steps $\Delta_{t}$ are chosen according to $\Delta_{t} \leq \eta
({\epsilon_{i}/\alpha_{i}})^{1/2}$, where $\epsilon_{i}$ is the
gravitational softening length of each particle, $\alpha_{i}$ is the
value of the local acceleration, and $\eta$ is a parameter that
specifies the size of the individual time steps and, consequently, the
time accuracy of the integration. The time integration conserved energy
to better than 0.15\% in all cases, which is adequate for the kind
of numerical simulations presented in this paper. The energy must be
conserved at such a level that the dynamics of the regions of interest
can be meaningfully probed.
We have also explicitly checked that our results are not compromised 
by choices of force-softening, time-stepping, or opening angle criteria 
in the treecode.

In what follows and unless otherwise explicitly stated,
we consider as our framework the concordance 
flat {\LCDM} cosmological model with present-day matter and vacuum 
densities $\Omega_{\rm m}=0.3$ and $\Omega_{\Lambda}=0.7$, respectively, 
dimensionless Hubble constant $h=0.7$, present-day fluctuation amplitude 
$\sigma_{8}=0.9$, and index of the primordial power spectrum $n=1.0$. 
Note also that in the remainder of this paper we use the terms ``satellites,'' 
``subhalos,'' and ``substructure'' indistinguishably.

\begin{table*}[tb]
\caption{Structural and numerical parameters of the satellite models}
\begin{center}
\begin{tabular}{ccccccc}
\tableline\tableline\\
\multicolumn{1}{c}{Model}&
\multicolumn{1}{c}{$N$}&
\multicolumn{1}{c}{$c$} &
\multicolumn{1}{c}{$r_{\rm s}$}  &
\multicolumn{1}{c}{$V_{\rm peak}$}  &
\multicolumn{1}{c}{$r_{\rm peak}$}  &
\multicolumn{1}{c}{$\epsilon_{\rm sat}$} 
\\
\multicolumn{1}{c}{}&
\multicolumn{1}{c}{}&
\multicolumn{1}{c}{}  &
\multicolumn{1}{c}{($h^{-1}\kpc$)} &
\multicolumn{1}{c}{($\kms$)} & 
\multicolumn{1}{c}{$(h^{-1}\kpc$)} &
\multicolumn{1}{c}{($h^{-1}\kpc$)}
\\
(1) &(2) &(3) &(4) &(5) &(6)&(7)
\\
\\
\tableline
\\
LR & $5 \times 10^{5}$ & $9$ & $5.4$ & $41.5$ & $11.7$ & $0.511$ \\
HR1 & $10^{7}$ & $21$ & $2.3$ & $51.3$ & $5.0$ & $0.021$ \\
HR2  & $10^{7}$ & $9$ & $5.4$ & $41.5$ & $11.7$  & $0.021$ \\
\\
\tableline
\end{tabular}
\tablecomments{
Col. (1): Satellite galaxy model. 
Col. (2): Number of particles. 
Col. (3): Concentration parameter.
Col. (4): Scale radius.
Col. (5): Maximum circular velocity.
Col. (6): Radius where the circular velocity peaks.
Col. (7): Softening length. Note that the virial mass of all satellite models
is $M_{\rm sat}=1.4 \times10^{10} \,h^{-1} \Mo$.
}
\end{center}
\label{table:model_parameters}
\end{table*}
%

\subsection{Substructure in a Fixed External Potential}
\label{subsection:fixed_pot}

In this section, we investigate the structural evolution of CDM substructure, 
using $N$-body simulations of satellites orbiting within the gravitational
potential of a static primary.
The satellites are modeled using the \citet[hereafter NFW]{navarro_etal96}
density profile, under the assumptions of spherical symmetry and isotropic
velocity dispersion tensors. These models are constructed using the procedure 
described in \citet{kazantzidis_etal04}, which produces self-consistent 
equilibria that do not suffer from numerical instabilities present in other
schemes, such as those that approximate the exact velocity distribution 
at any given point in space by a Gaussian.

The NFW density profile is given by
\beq
\rho(r)=\frac{\rho_{\rm s}} {(r/r_{\rm s}) (1+r/r_{\rm s})^2} 
\qquad\hbox{($r \leq r_{\rm vir}$)} \ ,
\label{equation:NFW_profile}
\eeq 
where the characteristic inner density $\rho_{\rm s}$ and scale radius 
$r_{\rm s}$, are sensitive to the epoch of halo formation and tightly
correlated with the halo virial parameters, via the concentration, 
$c\equiv r_{\rm vir}/r_{\rm s}$, and the virial overdensity, $\Delta_{\rm vir}$.
Since the NFW density profile corresponds to a 
cumulative mass distribution that diverges as $r\rightarrow\infty$,
we introduce an exponential cutoff for $r > r_{\rm vir}$. 
The latter sets in at the virial radius and turns off the profile
on a scale $r_{\rm decay}$, which is a free parameter and controls the
sharpness of the transition:
\beq
\rho(r)=\frac{\rho_{\rm s}} {c (1+c)^2} 
\left(\frac{r}{r_{\rm vir}}\right)^{\epsilon}
\exp\left(-\frac{r-r_{\rm vir}}{r_{\rm decay}}\right) \\
(r>r_{\rm vir}).
\label{equation:exp_cutoff}
\eeq
In order to ensure a smooth transition between equations (\ref{equation:NFW_profile})
and (\ref{equation:exp_cutoff}) at $r_{\rm vir}$, we require the logarithmic slope 
there to be continuous. This implies 
\beq 
\epsilon=-\frac{1+3c}{1+c} +
\frac{r_{\rm vir}}{r_{\rm decay}}.
\label{equation:eps}
\eeq 

The satellites' virial mass is equal to $M_{\rm sat}=1.4 \times10^{10} \,h^{-1} \Mo$,
corresponding to a circular velocity at the virial radius of $V_{\rm vir}= 35\kms$
for the adopted {\LCDM} model.
However, a dark matter halo of a given mass and size does not have a unique 
NFW profile. Indeed, the parameter that controls the shape of the 
density profile is the concentration $c$, and higher values of concentration 
correspond to a larger fraction of the virial mass being contained in the inner
regions of the halo. For a given mass, the measured scatter in this parameter 
reflects mainly the different formation epochs 
\citep{bullock_etal01a,eke_etal01,wechsler_etal02}.
For the adopted {\LCDM} model, the median concentration value 
for an object of this mass scale is 
$c_{\rm sat} = 21$, with the $2\sigma$ deviation given by
$c_{\rm sat} \sim 11-40$ \citep{bullock_etal01a}.

Here we present results for two high-resolution simulations 
($N=10^{7}$) of subhalos having concentration parameters equal 
to $c_{\rm sat}=21$ (HR1) and $c_{\rm sat}=9$ (HR2).
We consider these two significantly different values of the 
concentration to single out readily how the evolution of the satellites' 
structure depends on this parameter.
We note that values as small as $c_{\rm sat}=9$ for the mass scale of our satellites
are possible within models with a lower fluctuation amplitude 
$\sigma_8$. In order to demonstrate the need for such a high mass resolution,
we run a simulation using a lower resolution of $N=5 \times 10^{5}$ 
particles with the lower value of the concentration $c_{\rm sat}=9$ (LR). 
We emphasize that even the latter run uses a factor of $5$ more particles 
than the highest-resolution simulations of H03. 

For the low resolution run, we choose a spline softening
length of $\epsilon=0.511 h^{-1}\kpc$. This value corresponds 
to an equivalent Plummer softening length
equal to that of the highest resolution cosmological simulation, GA2, of S02.
For models HR1 and HR2, we choose a gravitational softening 
of $\epsilon=0.021 h^{-1}\kpc$, comfortably smaller than the typical sizes
of the dSph satellites of the Milky Way which constitute
the ultimate target of the present study.
Numerical and structural parameters of all models are
summarized in Table~\ref{table:model_parameters}.
For all the runs, we followed the time evolution of the density,
$\rho(r)$, and circular velocity, $V_{\rm c}(r)$, profiles
of the satellites. We explore below how these evolve
depending on the structural properties of the subhalos
and on the numerical resolution.

The orbits of the satellites are influenced only by the external, spherically 
symmetric static tidal field, which is represented 
by the logarithmic halo potential, 
\beq
\Phi = \sigma^{2}\ln(R^{2}+R_{\rm c}^{2}) \ ,
\label{equation:log_potential}
\eeq
where $\sigma$ is the one dimensional velocity dispersion
and $R_{\rm c}$ is the core radius. 
We use $\sigma=127.3\kms$ and $R_{\rm c}=0.7 h^{-1}\kpc$,
resulting in a circular velocity profile that flattens out already at $\sim 3\kpc$ and reaches
an asymptotic value of $v_{0}= \sqrt 2 \sigma \sim 180 \kms$.
This modeling gives a total mass 
within $100\kpc$ of the center of the fixed potential
equal to $M_{R<100\kpc} = 5.25 \times 10^{11} \,h^{-1} \Mo$.
This value is at the upper limit of estimates for the mass of the Milky Way
from satellite motions \citep{kochanek96} and dynamics of the Magellanic Clouds
\citep{lin_etal95}, and it is within the range of 
models for the Milky Way proposed in the context of {\LCDM} \citep{klypin_etal02}. 
This choice serves our primary aim, which is to investigate the change of the 
satellites' internal structure in a regime in which the tidal shocks are very strong. 
Note also that the total mass within the pericenter of the orbit is equal to 
$M_{R<25\kpc} = 1.3 \times 10^{11} \,h^{-1} \Mo$.

The orbits of the Milky Way dSphs are currently poorly constrained 
observationally. Nevertheless, their current distances, which give 
an indication of the apocenter of their orbits, coupled with studies of the orbital properties
of cosmological halos, can be used to constrain the orbital parameters of the satellites.
In particular, the model satellites are placed on an eccentric orbit 
with an apocenter radius of $r_{\rm apo}=105 h^{-1}\kpc$ and 
$(r_{\rm apo}/r_{\rm per})=\, (6/1)$, close to the median ratio of apocentric to
pericentric radii found in high resolution cosmological $N$-body
simulations \citep{ghigna_etal98,tormen_etal98}.
The center of the external potential is always set to be $(x,y,z) = (0,0,0)$.
We begin all our simulations by placing the satellites at apocenter,
and we integrate the orbits forward for $7$ Gyr.
This timescale corresponds to more than three 
orbital periods ($T_{\rm orb} \sim 2.25$ Gyr) 
in the chosen orbit and already represents a significant 
fraction of the cosmic time.
Our approach neglects the effects of the dynamical friction and
the response of the primary to the presence of the satellite.
However, this choice is justified given the difference in the mass (almost a factor of 50)
and size of the two systems and the anticipated rapid and substantial mass loss
owing to tides \citep[see also][]{taffoni_etal03}.

\begin{figure*}
\centerline{\epsfxsize=6in \epsffile{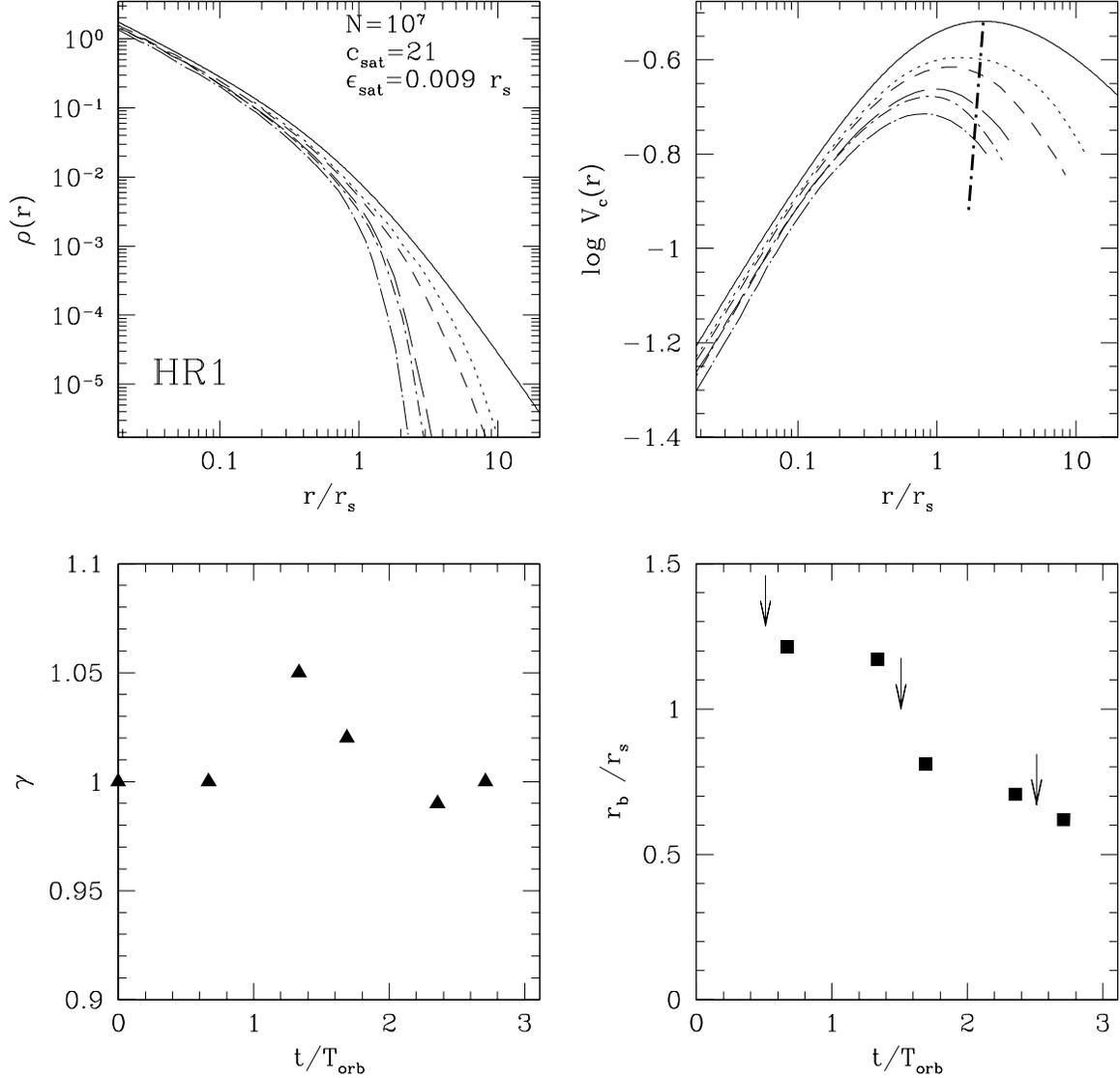}}
\caption{{\it Top:} Evolution of the density ({\it left})
and circular velocity ({\it right}) profiles of the bound mass 
for the high resolution model HR1.
The number of particles, $N$, the concentration parameter, 
$c_{\rm sat}$, and the softening length, $\epsilon$, are indicated  
({\it upper right-hand corner}). The scale radius of this model is  
$r_{\rm s}=2.3 h^{-1}\kpc$. The orbital evolution shown corresponds to approximately 
three orbital periods ($T_{\rm orb} \sim 2.25$ Gyr).
Profiles are shown at every odd quarter of the orbit 
between apocenter and pericenter, after allowing the satellites to pass 
the first pericenter, and are plotted from the force resolution ($2\epsilon$)
outward. The lines from top to the bottom, in order of decreasing central density,
show the profiles at $t= (0,0.67,1.33,1.69,2.36,2.71)\,T_{\rm orb}$.
The density and circular velocity are given in units of $M_{\rm sat} / r_{\rm s}^{3}$ and
$(G M_{\rm sat} / r_{\rm s})^{1 /2}$, respectively.
The thick dotted line indicates the expected relation between
$V_{\rm max}$ and $r_{\rm max}$ for field halos (see text for details).
{\it Bottom:} Evolution of the central density slope $\gamma$ ({\it left})
and break radius $r_{\rm b}$ ({\it right}) of the fitting formula (eq. [\ref{equation:fit}])
that describes the structure of our subhalos. The latter fit parameter is 
expressed in units of the scale radius, $r_{\rm s}$, 
of the initial model. Downward arrows indicate the pericentric passages.
The substructures maintain the same steep central density slope down to the limit 
of our force resolution as they get tidally stripped.
Tides, however, shift inward the break radius $r_{\rm b}$ of the fitting function.
Note the considerable change in $r_{\rm b}$ after each pericentric passage.
\label{fig1}}
\end{figure*}
\begin{figure*}
\centerline{\epsfxsize=6in \epsffile{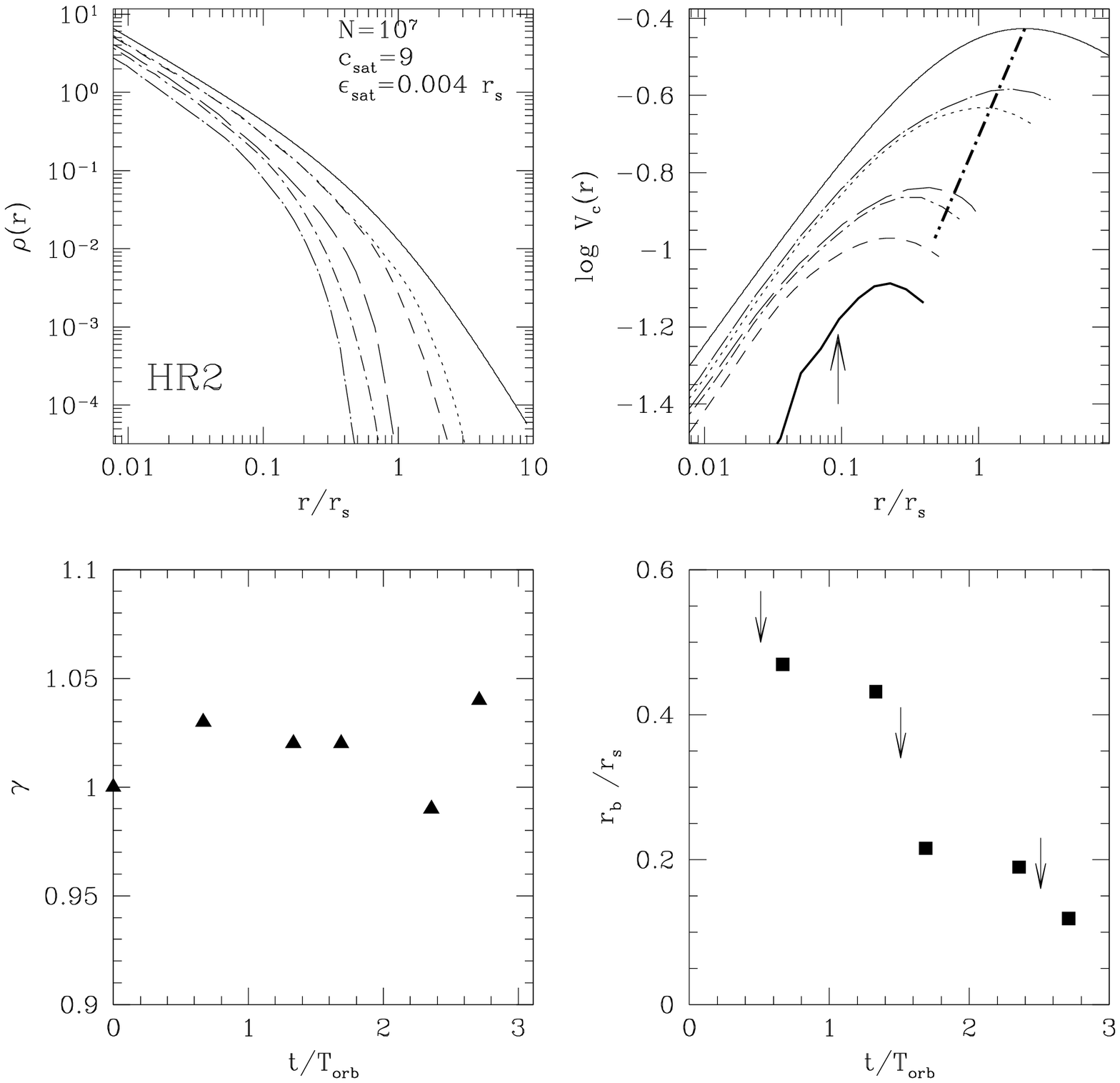}}
\caption{Same as Figure~\ref{fig1}, but for model HR2. 
The scale radius is $r_{\rm s}=5.4 h^{-1}\kpc$.
Compared to model HR1, the decrease in the central density and maximum circular 
velocity, $V_{\rm max}$, is significantly more pronounced, because of the higher
binding energy for more concentrated models.
Tidal interactions truncate this model at smaller physical radii than in model HR1, 
which is reflected in the evolution of the break radius 
({\it bottom right}).
The thick solid line shows the circular velocity profile of the low resolution 
satellite (LR) at $t=2.71\,T_{\rm orb}$. 
Even though this timescale is the same as the one
corresponding to the last curve in the high-resolution run, 
the low resolution satellite has lost
substantially more mass, and its central slope indicates a shallower
inner density profile. Upward arrow indicates the softening used in 
the low resolution simulation. 
\label{fig2}}
\end{figure*}

The tidal field alters the structure of the satellites
through a combination of strong 
tidal shocks occurring at each pericentric passage and
gradual mass loss throughout the orbital evolution.
In the top panels of Figures~\ref{fig1} and~\ref{fig2},
we present the evolution of the density ({\it left}) and circular velocity ({\it right}) 
profiles of the bound remnant of the initial models HR1 and HR2, respectively.
After allowing the satellites to pass past the pericenter for the first time, 
we show the profiles at every odd quarter of 
the orbit between apocenter and pericenter. These profiles are plotted from the
force resolution ($2\epsilon$) outward. The choice of how to determine the 
bound remnants' center is crucial for our intended analysis, since quantities such as 
the spherically averaged density and circular velocity profiles are quite sensitive 
to the center's definition. In this study, we identify the center adopting 
the most bound particle method, which agrees very well 
with the center of mass recursively calculated using spherical regions
of decreasing radius. Indeed, we confirmed that both methods give converging results
by comparing the resulting density profiles for some of our stripped satellites. 
This comparison yielded identical profiles for the remnants at all radii between our 
force resolution and the tidal radius for all the cases we tried.

We determine the mass that remains self-bound as a function of time 
using the following iterative scheme: in the rest frame of the most bound particle,
we calculate the binding energies of all the other particles, using the tree-based
gravity calculation performed by PKDGRAV, and we remove all those with positive 
binding energy. This calculation of binding energies
and subsequent removal of unbound particles is repeated until
no more particles can be removed or no bound remnant is found 
(i.e., all particles are removed).  In practice, this iterative procedure
converges rapidly and ensures that the true bound entity will be 
identified. Note that this technique is essentially the same one used by most
group finding algorithms, such as the publicly available SKID \citep{stadel01}, which
we use for the substructure in the cosmological simulations analyzed in 
\S~\ref{subsection:cosmo_sim}, where removing the background 
potential is more difficult, but it has the advantages of 
(1) using a tree structure for the potential calculation which
requires of the order of $O(N \log N)$ operations instead of
$O(N^2)$, where $N$ is the number of particles 
in the remnant, and (2) having a parallel implementation for very large
$N$. In this way, we can handle a much larger number of particles 
than would be possible with SKID, and at a fraction of the computational burden.

In the above analysis, we used equal-size logarithmic bins.
Different number of bins were used, depending on the stage of the orbital evolution, 
ensuring that in each case we have a sufficiently large number ($>1000$) of particles 
in each bin. This choice of binning simply minimizes the noise in the resulting profiles.
The central density of model HR2 decreases between the initial and final values
by almost a factor of $3$ more than the decrease in model HR1
over the same timescales. This is due to the fact that the initial concentrations are 
quite different. The concentration parameter of model HR1 is considerably higher than
that of HR2, and thus the former model is more resilient to tidal heating.
The same effect is clearly seen on the evolution of the circular velocity profiles.
In particular, the overall change in the $V_{\rm max}$ is significantly 
more pronounced in model HR2 than in model HR1 for the same 
timescales. The thick dotted lines in the top right panels 
of Figures~\ref{fig1} and~\ref{fig2} 
indicate the expected relation between $V_{\rm max}$ and the 
distance at which the maximum circular velocity,
$r_{\rm max}$, occurs for {\it field halos} \citep{colin_etal03}. 
The latter is a measure of the concentration of the satellites.
Tidal stripping moves the subhalos to the left of the expected relation,
so that for a given $V_{\rm max}$ these systems have smaller 
$r_{\rm max}$ and hence higher effective concentrations
than field halos. For example, after $\sim 6$ Gyr 
the $V_{\rm max}$ and $r_{\rm max}$ of the $c_{\rm sat}=9$ 
satellite decrease by a factor of $3.5$ and $8.5$, 
respectively. For the same change of $V_{\rm max}$, however, one would
expect $r_{\rm max}$ to vary by just a factor of $3$ from the relation for field halos.
The higher relative densities of heavily stripped
halos may account for the fact that Draco, being considerably closer to the
Milky Way, has a mass-to-light ratio higher than that of Fornax, despite their 
similar stellar velocity dispersions. 

\begin{figure*}
\centerline{\epsfxsize=6in \epsffile{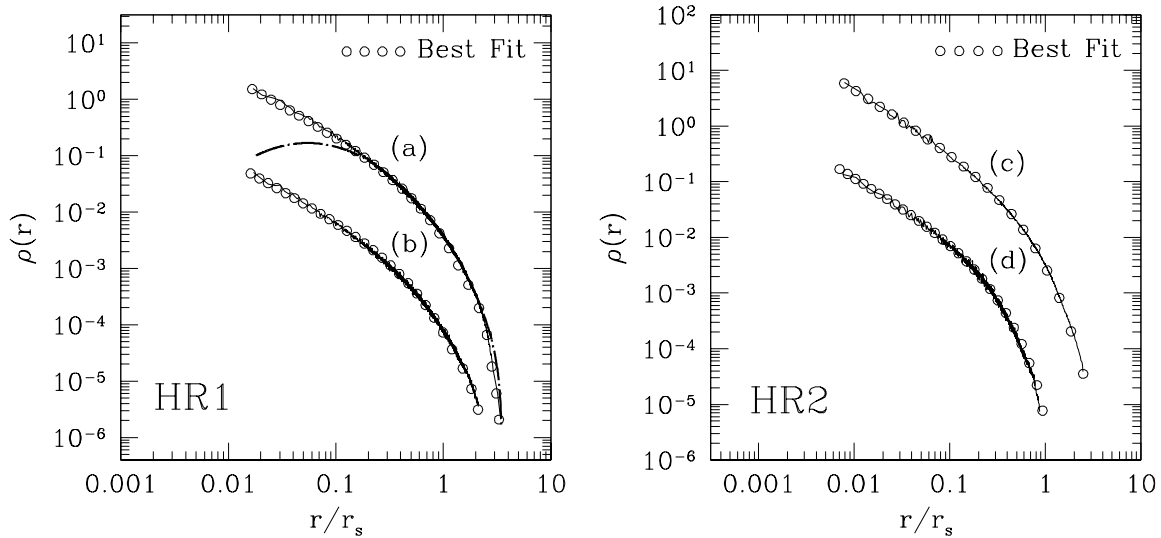}}
\caption{Density profiles of substructure halos from our individual
satellite simulations ({\it solid curves}),
together with fits to the density structure obtained using 
eq. (\ref{equation:fit}) ({\it circles}).
The profiles are shown for models HR1 ({\it left}) and 
HR2 ({\it right}) and for two different timescales in the orbital evolution.
In both panels, the density profiles corresponding to the lower curves 
are vertically shifted downward by $1.5$ dex for clarity.
Curves (a) and (b) correspond to $ t=1.69\,T_{\rm orb}$ and 
$t=2.71\,T_{\rm orb}$, respectively. These curves are best described by 
fitting parameters  
$(\gamma, r_{\rm b} / r_{\rm s})=(1.02, 2.69)$ and 
$(\gamma, r_{\rm b} / r_{\rm s})= (1.00, 2.06)$, respectively.
Curves (c) and (d) correspond to $ t=1.33\,T_{\rm orb}$ and 
$t=2.36\,T_{\rm orb}$ and the best fit parameters are 
$(\gamma, r_{\rm b} / r_{\rm s})= (1.02, 0.43)$ and 
$(\gamma, r_{\rm b} / r_{\rm s})= (0.99, 0.19)$, respectively.
The fits are satisfactory over $4$ orders of magnitude in density 
and $2$ orders of magnitude in radius and for satellites
having concentration parameters that differ by more 
than a factor of $2$. The thick dotted line shows the density 
profile corresponding to the parabolic circular velocity fitting formula 
proposed by S02 with $a=0.45$, the median value of their best fits 
to substructure halos (see text for details).
\label{fig3}}
\end{figure*}

The density structure of our subhalos can be described
through the following simple formula 
\beq    
\rho(r)\propto r^{-\gamma} \exp ({-\frac{r}{r_{\rm b}}})\ ,
\label{equation:fit}
\eeq
where $\gamma$ denotes the central slope of the substructure density 
profile and $r_{\rm b}$ an effective ``break'' radius describing the outer cutoff 
imposed by the tides. In the bottom panels of Figures~\ref{fig1} and~\ref{fig2}, 
we show the evolution of these two fitting parameters. Note that
$r_{\rm b}$ is expressed in units of the scale radius, $r_{\rm s}$, of the initial models. 
The robustness of the aforementioned binning procedure was verified
by comparing these results against those when the particles were binned
in spherical shells with the same number of particles in each bin.
In this case, we choose as bin center the average radius of all particles in 
the given bin, and we assign equal statistical weight to each radial bin.
The two procedures yielded almost identical
values for $\gamma$ and $r_{\rm b}$ for the entire orbital evolution. 
Note that in the bottom panels of Figures~\ref{fig1} 
and~\ref{fig2}, we present our results adopting the latter binning procedure,
and the same is also true for results shown in Figure~\ref{fig3}.

The bottom left panels of Figures~\ref{fig1} and~\ref{fig2}
demonstrate that tidal effects do not serve to 
reduce the central density cusp down to the limit of our force
resolution. This is a fundamental result of this paper that is independent
of the concentration of the satellites and valid for their entire orbital
evolution. Most of the tidally stripped mass is removed from the outer parts
of the subhalos, steepening  the outer density profiles
and shifting the break radius $r_{\rm b}$ inward. This just reflects 
the fact that tides truncate the satellite models at increasingly 
smaller physical radii. The two bottom right panels illustrate that the lower concentration
satellite (HR2) gets truncated at smaller radii than its higher concentration 
counterpart. In both models, the break radii experience the biggest decrease
after each pericentric passage, but this decrease is more
pronounced in model HR2.
Between pericenters the break radii decrease in a smoother manner.

Even though this result is also verified in the lower resolution model LR 
when we compared the density profiles, it is interesting to note that the same 
is not true for the circular velocity profiles.
Indeed, convergence in the latter profile is significantly more problematic, and this 
is highlighted in the top right panel of Figure~\ref{fig2}, where
the circular velocity profile of the low resolution satellite 
is shown for comparison at $t=2.71\,T_{\rm orb}$ ({\it thick solid line}).
The downward arrow indicates the softening used in this simulation.
The circular velocity profile of a model satellite having
a factor of $20$ fewer particles than its high-resolution counterpart
and evolved on the same orbit is significantly lower and has a steeper inner gradient.
This is entirely due to resolution effects.
The fact that the convergence in the circular velocity is harder to attain
is not surprising, since it is a cumulative quantity 
(see also below,\S~\ref{subsection:cosmo_sim}).
This warns against using the circular velocity to compute the structural 
properties of satellites, as done in S02.

Figures~\ref{fig1} and~\ref{fig2} indicate that the inner satellite
regions corresponding to sub-kiloparsec scales 
are much less affected by the tides than the outer ones. 
In particular, the satellite with $c_{\rm sat}=21$ suffers almost
no change, even in the value of the central density. 
Subhalo particles that have orbital times shorter than the 
shock duration will be only marginally affected by the shock itself. 
This is known as adiabatic correction \citep{weinberg95}.
Especially for the satellite with the highest central density ($c_{\rm sat}=21$),
the adiabatic correction to the impulse approximation
is extremely important, as it reduces significantly the 
predicted amount of shock heating. We can calculate the variation
of the kinetic energy of a particle located at some distance $r$
from the center of the satellite, $\Delta E$, due to
impulsive heating at each pericentric passage in the extended mass
distribution of a primary isothermal halo \citep[see also][]{mayer_etal02,taffoni_etal03}.
\citet{gnedin_etal99} have shown that for orbits as eccentric 
as those considered here, we can approximate the trajectory of 
the satellite with a straight line path and write
\beq
<\Delta E>=\frac{1}{2}\left(\frac{GM_0}{R_{\rm per}^{2}V_{\rm per}}\right)^{2}
\frac{r^{2}\pi^{2}}{3}\left(\frac{R_{\rm per}}{R_{\rm max}}\right)^{2} 
(1 + \omega\tau)^{-2.5} \ ,
\label{equation:adiab_correction}
\eeq
where $M_0$ and $R_{\rm max}$ are the total mass and radius, respectively,
of the primary halo, which are assumed equal to the mass and radius 
within the apocenter of the satellite's orbit, and $V_{\rm per}$ is the satellite's velocity
at the pericenter of the orbit. These parameters are equal to 
$M_0=8.4 \times 10^{11} \,h^{-1} \Mo$, $R_{\rm max}=105 h^{-1}\kpc$, and 
$V_{\rm per}=345\kms$. The last term in the product is the first-order adiabatic 
correction with $\omega$ and $\tau$ being the orbital frequency for a particle at 
radius $r$ and the duration of the tidal shock, respectively.

For a particle at a distance of $r=0.7 h^{-1}\kpc$ from the
center of the satellite, $\Delta E$ without the adiabatic 
correction is roughly equal to 20\% of 
its binding energy, $E_{\rm bind}$, after three tidal shocks.
Instead, $\Delta E/E_{\rm bind}$ is already greater than $1$ at a 
radius of $r=1.75 h^{-1}\kpc$, despite the fact that the tidal radius of the satellite is 
much larger. However, once the adiabatic correction is included, 
the ratio $\Delta E/E_{\rm bind}$ is reduced by nearly 90\%, explaining 
why the satellite is barely affected by the
tidal shocks in its inner regions. Note that in order to incorporate 
the effect of the adiabatic correction in the simulations,
one needs to model the inner regions of the subhalo 
quite accurately. This means resolving down to sub-kiloparsec scales in the first 
place and also avoiding non-equilibrium initial conditions or coarse
mass resolution. The latter could affect the orbital
frequencies of the dark matter particles, which might well end up in
a region of phase space that is not protected by the adiabatic invariance
anymore (this problem is reminiscent of the difficulty that $N$-body simulations
of low resolution have in resolving resonances in stellar 
systems; see \citet{weinberg_katz02}).

Figure~\ref{fig3} shows the density profiles for two different 
timescales in the orbital evolution of models HR1 and HR2 (solid curves).
Fits to the density structure of the substructure halos 
using equation (\ref{equation:fit}) and obtained by $\chi^2$ minimization 
are also shown, as circles.
Model HR1 ({\it left}) is plotted at $ t=1.69\,T_{\rm orb}$
and  $2.71\,T_{\rm orb}$, which correspond to 
$\sim 0.4$ Gyr after the satellite has concluded its second
and third pericentric passages, respectively.
Model HR2 ({\it right}) is plotted at $ t=1.33\,T_{\rm orb}$ 
and $2.36\,T_{\rm orb}$, which correspond to 
$\sim 0.4$ Gyr before the conclusion of the satellite's second
and third pericentric passages, respectively.
Note that for clarity we offset the lower curves in both panels
by $1.5$ dex. Clearly, the structure of our stripped satellites
is reasonably well reproduced by the proposed fitting formula.
We should note that this formula
works well in the case of the lower concentration satellite 
for the entire orbital evolution. On the other hand, the case of the $c_{\rm sat}=21$ 
satellite is somehow slightly different. The fitting function works better for the 
last stages of the orbital evolution, at which the satellite
has suffered significant mass loss.
This is just reflecting the resilience to tidal stripping
of the high concentration satellite during the early stages 
of the evolution and the fact that the mass loss is more gradual
compared to the low concentration model, even from the outer regions.
Finally, for comparison we show the density 
profile resulting from the fitting formula proposed by S02 ({\it thick dotted line})
to describe the structure of their subhalos (see their eq. [1]),
plotted down to the limit of our force resolution. 
The values for $r_{\rm max}$ and 
$V_{\rm max}$ in their formula are taken directly from the particular 
subhalo. In addition, we adopt a value for the parameter $a$ in their  
equation that is equal to the median of their best fits, $a=0.45$.
Compared to the density profiles describing the 
structure of our satellites, the latter profile has a substantially shallower slope 
on scales comparable to the sizes of the dSphs.
Interestingly, the density reaches even negative values at a finite radius,
demonstrating that its use should be avoided.

\subsection{Substructure in Hierarchical Cosmological Simulations}
\label{subsection:cosmo_sim}
%
\begin{figure*}
\centerline{\epsfxsize=6in \epsffile{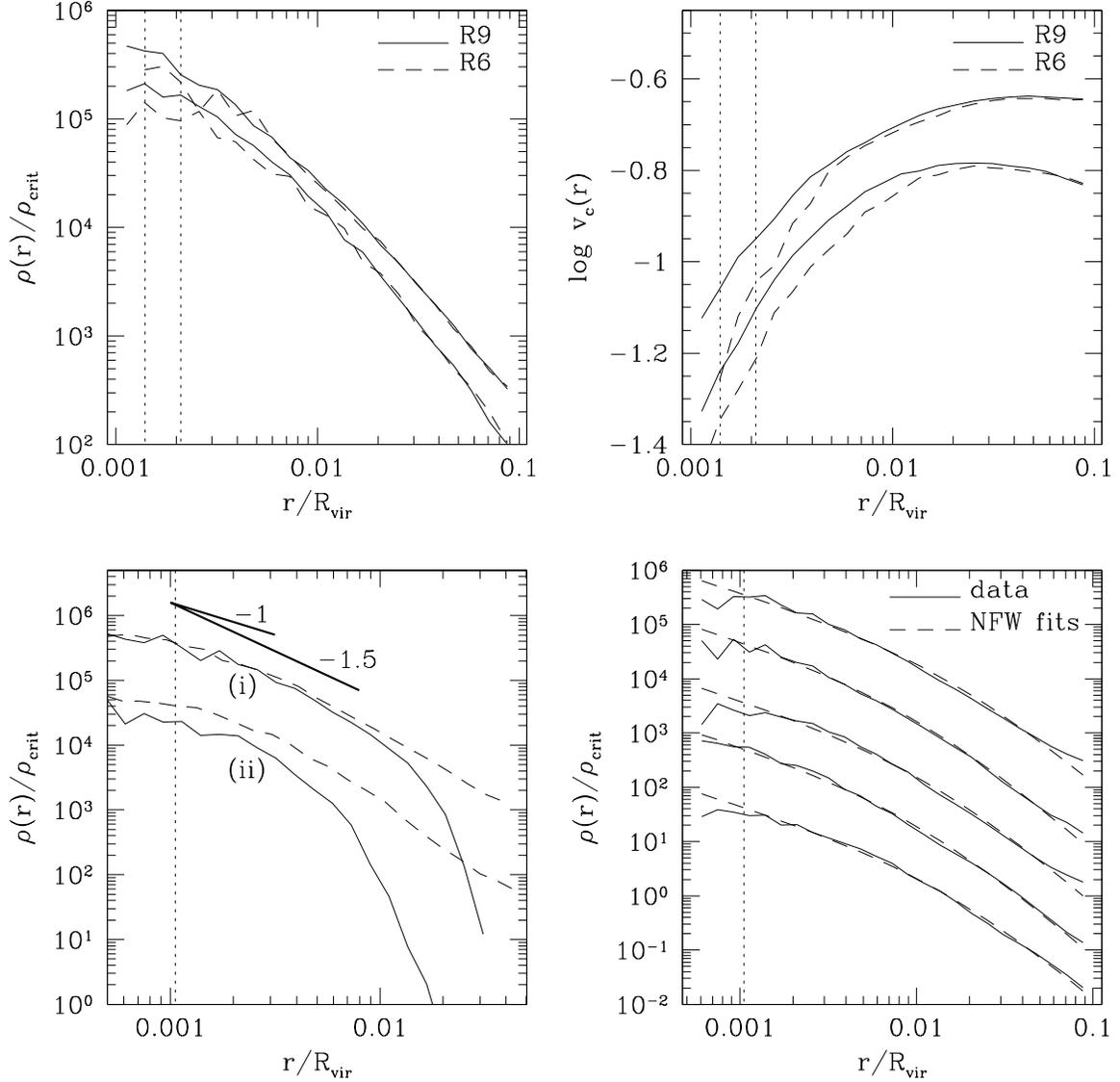}}
\caption{{\it Top:} Density ({\it left}) and circular
velocity ({\it right}) profiles of two subhalos simulated at 
two different resolutions. Circular velocities are expressed
in units of the maximum circular velocity of the parent halo.
In simulation R9 (solid lines), the subhalos contain $3.4 \times10^4$ and $1.4 \times10^4$ 
particles, while in R6 (dashed lines) they contain a factor of $3.375$ fewer particles.
The convergence in the circular velocity profiles between the two resolutions 
is significantly slower on account of the cumulative nature of this quantity. 
This demonstrates that it is erroneous to use the circular velocity to compute 
the structural properties of substructure.
{\it Bottom left:} Density profiles of two heavily stripped 
substructure halos in simulation R12 before entering 
the primary halo ({\it dashed lines}) and 
at present ({\it solid lines}). Subhalo (i) is shown at $z=0.80$ and just before
the second pericentric passage. Subhalo (ii) is shown at $z=0.94$ and
just before the fourth pericentric passage and is offset by $1$ dex to avoid overlap.
The numbers near the thick solid lines indicate the power slope of those lines.
Both subhalos have of the order of $N=2 \times10^4$ particles before entering the host.
The cosmological subhalos maintain their steep inner density slope even
after several pericentric passages.  
{\it Bottom right:} Density profiles of the five most massive halos in simulation R12
(solid lines). The best-fit NFW profiles are also plotted ({\it dashed lines}).
To avoid overlap the lower four density profiles are vertically shifted
by $1$, $2$, $3$, and $4$ dex from the top to the bottom.
These five halos are resolved with more than $2 \times10^5$ particles. 
In all panels the vertical dotted lines show the adopted force resolution.
\label{fig4}}
\end{figure*}
In this section, we present results from a set of high resolution
{\LCDM} cluster simulations \citep{diemand_etal04b}.
The initial conditions are generated with
the GRAFIC2 package \citep{bertschinger01}. We begin with a 
$300^3$ particle cubic grid with
a comoving cube size of $300\Mpc$ (particle mass 
$m_{\rm p} = 2.6  \times 10^{10} \,h^{-1} \Mo$). At $z=0$ we identify
several clusters for resimulating at higher resolution, using refinement
factors of $6$, $9$, and $12$ in length ($216$, $729$, and $1728$ in mass), 
so that the mass resolution is $m_{\rm p} = 1.5 \times 10^7 \,h^{-1} \Mo$ in the highest 
resolution run. We label these three runs as R6, R9 and R12, after their refinement 
factors. At $z = 0$ the refined cluster contains $1.8\times 10^6$ particles within the virial 
radius in run R6, $6\times 10^6$ in R9, and $1.4\times 10^7$ in R12.
Note that we define the virial radius of the clusters, $R_{\rm vir}$, to be the distance
from the center at which the mean enclosed density is 
$178 \Omega_{\rm m}^{0.45} \approx 103.5$ times the critical value $\rho_{\rm crit}$
\citep{eke_etal96}. The softening length is comoving from the start of the simulation 
($z \simeq 40$) to $z = 9$. From $z = 9$ until the present we use a physical softening
length of $2.1$, $1.4$, and $1.05 \times 10^{-3} R_{\rm vir}$ for R6,R9, and R12 respectively.

In the top panels of Figure~\ref{fig4}, we show the density ({\it left}) 
and circular velocity ({\it right}) profiles of two distinct substructure halos
simulated at two different resolutions. In simulation R9 ({\it solid lines}), the two subhalos 
contain $N=3.4 \times10^4$ and $N=1.4 \times10^4$ 
particles, respectively, whereas in simulation R6 ({\it dashed lines})
the same subhalos contain a factor of $3.375$ fewer particles.
The vertical dotted lines show the adopted force resolution.
Densities are expressed in units of the present critical density for closure,
$\rho_{\rm crit}=3H_{0}^2/8\pi G$, and the virial radius 
of the parent halo is equal to $R_{\rm vir}=1.225 h^{-1}\Mpc$.
The profiles of the cosmological satellites are to be trusted only up to
the resolution limit set by two body relaxation \citep{diemand_etal04a}.
We note that the convergence of the circular velocity profiles is much weaker
than that of the density profiles at different resolutions, in agreement with 
the results presented in \S~\ref{subsection:fixed_pot}.
As the circular velocity is a cumulative quantity, numerical effects show up at 
comparatively larger radii, where the density profiles have already converged, biasing the 
circular velocity profiles.
Therefore, one should be especially cautious when using 
the circular velocity profiles obtained at one particular
resolution to provide fitting functions to systems whose typical 
scales fall below such resolution, as done in S02.

In the bottom panels of Figure~\ref{fig4}, we present results
regarding substructure halos taken from our highest
resolution simulation (R12).
The density profiles of two heavily stripped subhalos before entering the primary halo 
({\it dashed lines}) and at present ({\it solid lines}) are shown in the bottom 
left panel. These satellites have of the order of $N=2 \times10^4$ particles before 
entering the host. The subhalo denoted ``(i)'' is shown at $z_{\rm in}=0.80$
and at present just before the second pericentric passage 
($r_{\rm min}= 0.21 R_{\rm vir}$). The second subhalo, denoted ``(ii),'' is 
offset by $1$ dex to avoid overlap and is shown at $z_{\rm in}=0.94$ and at present
just before the {\it fourth} pericentric passage ($r_{\rm min}= 0.15 R_{\rm vir}$).
The thick solid lines indicate two critical values for the inner slope,
$\rho(r)\propto r^{-1}$ and $\rho(r)\propto r^{-1.5}$.
This plot illustrates that the substructures maintain the steep inner density slope
that they had before entering the primary halo, even after several 
pericentric passages. These heavily stripped satellites have outer profiles consistent 
with an exponential cutoff. Even though these cosmological simulations are 
state-of-the art by the current standards, the resolution in these stripped satellites
is still too low for us to draw robust conclusions regarding the evolution of
their inner structure.

Finally, the bottom right panel of Figure~\ref{fig4} shows the
density profiles of the five most massive substructure halos in this simulation 
({\it solid lines}) together with the best fit NFW profiles ({\it dashed lines}).
To avoid overlap, the lower four density profiles are each 
vertically shifted by $1$ dex from each other.
These five halos are resolved with more than $N=2 \times10^5$ particles,
and their density profiles are described reasonably well by the NFW profile.
These halos entered the host system quite late and have typically made only
one orbit; however, these most massive halos represent the ones that would be
associated with the dSphs, according to S02.
We can plausibly trust their central structure to $0.002R_{\rm vir}$, which 
corresponds to about $500\pc$ when scaled to the Galaxy. 
At this radius the density profiles are still cuspy, with central slopes between 
$-1$ and $-1.5$.

We end by emphasizing that the behavior of the satellites in the time-dependent 
cosmological tidal field is not obviously similar to that our individual 
subhalos exhibit when evolved in a static host potential.
The latter are spherical systems with isotropic velocity dispersion tensors, whereas
the cosmological satellites presented here are in general triaxial, anisotropic systems,
and they suffer additional artificial heating from background particles and
physical heating from encounters with other substructures.
It is remarkable, therefore, that our findings regarding the way that tidal 
interactions affect the central density cusps
converge in these two radically different regimes: the inner density profiles 
remain cuspy to the resolution limit of the simulations.

\section{DWARF SPHEROIDAL KINEMATICS IN CDM SUBHALOS}
\label{section:dSph_kinematics}
%
\begin{figure*}
\centerline{\epsfxsize=3.8in \epsffile{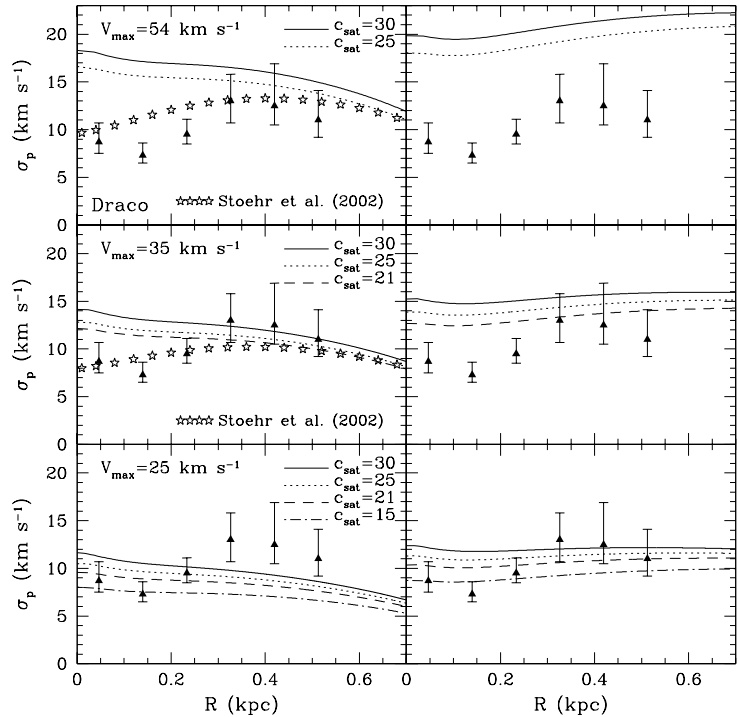}}
\caption{Observed line-of-sight velocity dispersions for the dSph 
Draco ({\it triangles}), compared with those predicted for 
stellar systems embedded in dark matter halos with structure similar to the ones
of our simulated subhalos.
Each pair of panels refers to a single value of $V_{\rm max}$, while
the various lines correspond to different values of
$R_{\rm max}$ in the same range as in S02 (see text for details).
The three plots on the left show results assuming a tidal radius for Draco
equal to the {\it optical radius} measured by \citet{odenkirchen_etal01} and also
adopted by S02. The panels on the right show results for the same range
of parameters, but for tidal radii $3$ times as large.
The filled points correspond to the results obtained assuming
the parabolic fits to the circular velocity profiles adopted by S02.
From top to bottom, the results for $V_{\rm max}=54$, $35$, 
and $25\kms$ are shown.
Globally, the observed velocity dispersion profile is better 
reproduced by subhalos with $V_{\rm max}=25\kms$ and assuming 
a significantly larger than the nominal tidal radius.
\label{fig5}}
\end{figure*}
Cosmological numerical simulations confirm that isolated halos in the mass range
$10^{7}-10^{10}\Mo$ have central cusps not shallower than $\rho(r) \propto r^{-1}$, on
scales comparable to the ones probed by the stars in the dSph satellites
\citep{moore_etal01,colin_etal03}.
Our results in \S~\ref{section:num_sim} indicate that an initial cuspy 
satellite will retain its cusp in the presence of a tidal field. 
Therefore, it is an interesting exercise to attempt
to reproduce the observed kinematics of the dSphs assuming cuspy dark matter 
distributions, and to constrain the maximum mass or circular velocity of their
host halos. The advantage of our approach lies in the fact that we have 
sufficient numerical resolution, and therefore there is no need of extrapolating 
the inner density slopes to scales smaller than the force resolution, as done by S02.

The dynamics of a spherically symmetric stellar distribution 
embedded in a spherical dark matter potential can be described 
by the lowest order Jeans equation, 
assuming that the two components are in equilibrium and 
that there are no net streaming motions (e.g., rotation): 
\beq    
\frac{\rm d}{{\rm d} r}  (\rho \sigma_{\rm r}^2) + \frac{2 \beta}{r} \rho 
\sigma_{\rm r}^2 + \rho \frac{{\rm d} \Phi}{{\rm d} r}=0 
\label{equation:Jeans}
\eeq
\citep{binney_tremaine87}, where $\rho(r)$ and $\sigma_{\rm r}(r)$ are the density 
profile and the radial velocity dispersion of the tracer stellar population, respectively,
$\Phi(r)$ is the underlying dark matter gravitational potential, and
$\beta$ describes the velocity anisotropy of the stars. The solution of the 
Jeans equation (\ref{equation:Jeans}) with the boundary condition
$\rho(r) \rightarrow 0$ at $r \rightarrow r_{\rm t}$ for $\beta$=const reads
\beq    
\rho \sigma_{\rm r}^2 = r^{-2 \beta}
        \int_{r}^{r_{\rm t}} r^{2 \beta} \rho \frac{{\rm d} \Phi}{{\rm d} r} \ {\rm d}r \ ,
\label{equation:Jeans_solution}
\eeq
where $r_{\rm t}$ is the tidal radius of the stellar component.

However, the quantity that can be measured and compared with observations
is the line-of-sight velocity dispersion, $\sigma_{\rm p}$, at projected distance $R$ 
from the center of the dwarf, which is given by
\beq
    \sigma_{\rm p}^2 (R) = \frac{2}{I(R)} \int_{R}^{r_{\rm t}}
    \left(1-\beta \frac{R^2}{r^2} \right) \frac{\rho \,
    \sigma_{\rm r}^2 (r, \beta) \,r}{\sqrt{r^2 - R^2}} \,{\rm d} r 
\label{equation:projected_disp}
\eeq
\citep{binney_mamon82,binney_tremaine87}, where $I(R)$ is the surface distribution 
of the tracer population:
\beq
    I(R) =
    2\, \int_{R}^{r_{\rm t}}
    \frac{r \,\rho(r)}{\sqrt{r^2 - R^2}} \,{\rm d} r.
\label{equation:surface_distr}
\eeq

We solve equation (\ref{equation:projected_disp}) using the customary 
King profile for the stellar population of the dSph galaxies,
\beq
\rho_{\star}(r) \propto \frac{1}{z^2} \left[\frac{1}{z}\cos^{-1}(z) - (1-z^2)^{1/2}\right]  
\label{equation:king}
\eeq
\citep{king62}, where $z \equiv [1 + (r/r_{\rm c})^2]/[1 + (r_{\rm t}/r_{\rm c})^2]$
and $r_{\rm c}$ denotes the core radius of the profile.
For Fornax and Draco, we use the parameters from \citet{mateo98} and 
\citet{odenkirchen_etal01}, respectively, similarly as in S02.
Note that the normalization in equation (\ref{equation:king}) is not important 
for the purpose of our analysis, since it cancels out.
Motivated by the structure of our simulated subhalos from the cosmological
simulations, we assume that the dark matter halos associated with the dwarf galaxies
follow the NFW density profile.
This provides a lower limit to the recovered 
velocity dispersions, since there is evidence in favor of steeper 
central density cusps on these scales \citep[e.g.,][]{moore_etal01}.
Our results are not sensitive to the presence of the exponential cutoff,
since the fitting formula of equation (\ref{equation:fit}) 
differs appreciably from the NFW density profile only near the break 
radius $r_{\rm b}$. This is larger than the tidal radii of the stellar 
components of the dSphs, where the integral in equation (\ref{equation:projected_disp}) 
has to be truncated anyway. In addition, we checked that any eventual differences in
the resulting projected stellar velocity dispersions were
negligible even when we considered models with stellar tidal radii significantly
larger than the nominal values (see below).

\begin{figure*}
\centerline{\epsfxsize=4.1in \epsffile{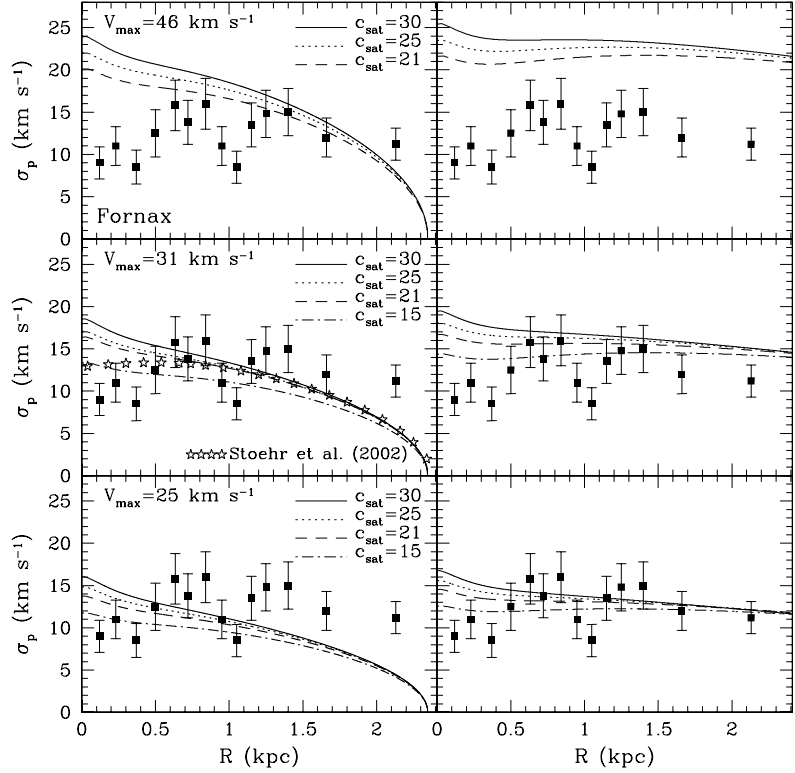}}
\caption{Same as Figure~\ref{fig5}, but for the dSph Fornax 
({\it squares}).
The three plots on the left show results assuming a value for the tidal radius 
taken from the review of \citet{mateo98}.
From top to bottom, the results for $V_{\rm max}=46$, $31$, 
and $25\kms$ are shown. Similarly to the case of Draco, the stellar
velocity dispersion is better reproduced by subhalos 
with $V_{\rm max}=25\kms$ and assuming a tidal radius $3$ times 
larger than the nominal value ({\it right}).
\label{fig6}}
\end{figure*}

We consider halos having both a maximum circular velocity, 
$V_{\rm max}$, and a corresponding radius, $R_{\rm max}=R(V_{\rm max})$, 
within the range considered by S02. 
In practice, for each value of $V_{\rm max}$ there is only
a limited range of halo concentrations such that $R_{\rm max}$ is consistent
with the values measured by S02 for their subhalos.
In Figures~\ref{fig5} and~\ref{fig6},
we show the results of the comparison for Draco and Fornax, respectively, 
assuming isotropic stellar orbits.
The observed velocity dispersion profiles are reproduced from 
\citet{mateo98} for Fornax and from \citet{kleyna_etal02} for Draco.
\begin{figure*}
\centerline{\epsfxsize=3.8in \epsffile{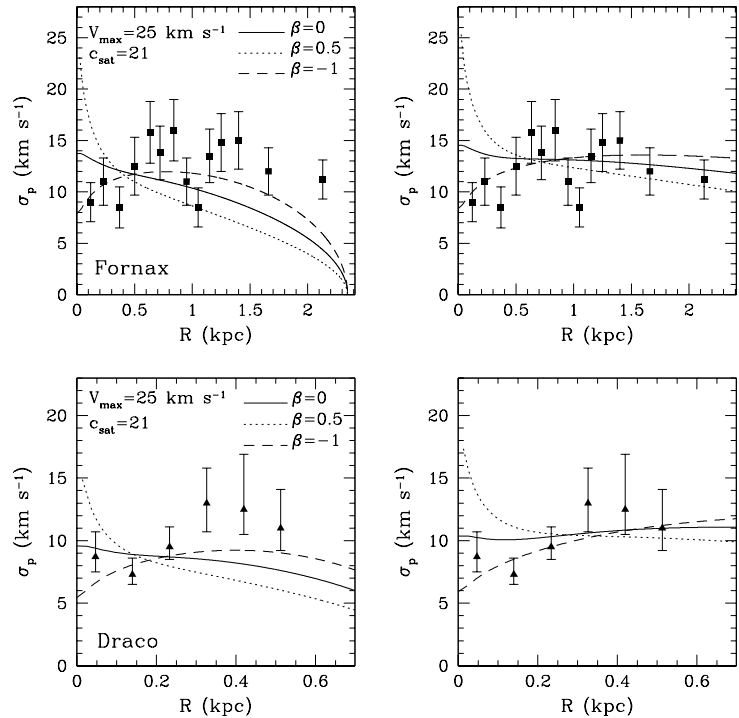}}
\caption{Kinematics of Fornax ({\it top}) and Draco ({\it bottom}), 
compared with those expected for stellar systems embedded within NFW subhalos 
with $V_{\rm max}=25\kms$. Different values for the anisotropy parameter $\beta$
in the velocity distribution of the stellar component have been considered.
The solid lines correspond to isotropic models, whereas the dotted and dashed
lines to radially and tangentially anisotropic models, respectively.
The plots on the left show results for values
of the tidal radii of the stellar component equivalent to those used by S02, 
while the plots on the right correspond to tidal radii $3$ times as large.
Models with mildly tangentially anisotropic distribution of stellar orbits better
reproduce the observed velocity dispersion profiles of both Draco and Fornax.
Radially anisotropic models overestimate the central stellar velocity dispersion
and make the curves decrease more steeply with distance, 
contrary to the trend in the observational data. 
\label{fig7}}
\end{figure*}
\begin{figure*}
\centerline{\epsfxsize=3.8in \epsffile{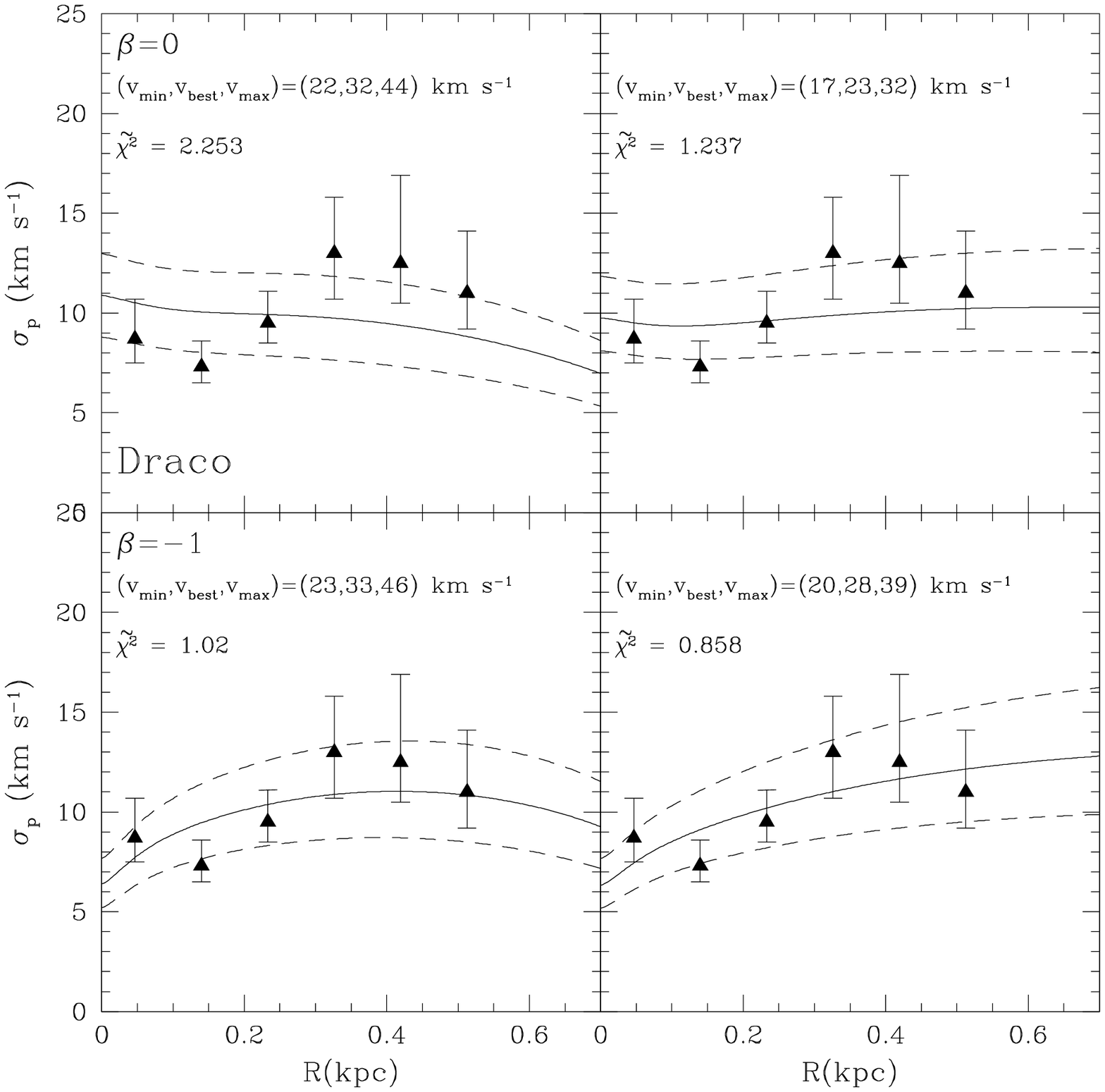}}
\caption{Best fit models ({\it solid lines}) and  $3\sigma$ intervals 
({\it dashed lines}) for the kinematics of Draco, assuming isotropic 
({\it top}) and tangentially anisotropic ({\it bottom}) models for the stellar distribution.
The stars are embedded in the potential wells of our stripped satellites
and follow a King profile.
The panels on the left show results adopting a tidal radius equal to
the nominal value, whereas the panels on the right show results 
for tidal radii $3$ times as large (see text for details).
Keeping the concentration parameter fixed ($c=21$), we fit by $\chi^2$ minimization the 
maximum circular velocity $V_{\rm max}$. 
The best fit value of $V_{\rm max}$ corresponds
to $\rm {v_{\rm best}}$, whereas $\rm {v_{\rm min}}$ and $\rm {v_{\rm max}}$ bracket 
the $3\sigma$ intervals.
The quality of the fits in terms of the reduced $\chi^2$ is also indicated for the 
different cases.
Subhalos with $V_{\rm max}$ in the range $20-35 \kms$ provide the best fit to the data, 
while $V_{\rm max} > 40 \kms$ are  $3\sigma$ or more away from the best fit.
\label{fig8}}
\end{figure*}
In some of these panels, where the relative information is available,
we overplot the results obtained adopting the parabolic fitting
function of the form proposed by S02 for their subhalos ({\it filled points}).
We remind the reader that the density profiles corresponding to this  
parabolic function have substantially shallower inner slopes at scales comparable 
to the sizes of the dSphs than the profiles describing the density structure
of our satellites. Each pair of panels refers to a single value of $V_{\rm max}$. 
The three panels on the left show results assuming a tidal radius equal to
that adopted by S02, while the panels on the right assume tidal radii $3$ 
times as large.

It is immediately apparent that subhalos with $V_{\rm max} \sim 50\kms$ 
would yield velocity dispersions significantly above the 
observed values for both Fornax and Draco, once the cuspy density 
profiles consistent with the structure of our satellites were used. 
The disrepancy becomes even worse 
if we allow the stellar component to have a tidal radius larger than the nominal 
value estimated by fitting the star counts to a King profile.
Recent observational studies of Draco show no evidence of tidal 
tails, lending support to the view that the optical radius
is only a lower limit to the true physical boundary of the system 
\citep{odenkirchen_etal01,kleyna_etal02,piatek_etal02,klessen_etal03},
and this might also be true for the majority of the dSphs.
By considering subhalos with $V_{\rm max} \sim 30\kms$, we match 
satisfactorily only the velocity dispersion data points at the outermost radii, 
which have the largest error bars. 
Subhalos with $V_{\rm max}=25\kms$ reproduce the observed stellar
velocity dispersions reasonably well for the entire range of parameters
considered. This conclusion is in disagreement with that reached by S02,
simply because of their use of a parabolic fitting function,
and suggests that there is no room for entirely solving the substructure
problem by simply changing the way the velocities are mapped.
Globally, the observed velocity dispersion profile is better 
reproduced by assuming a larger tidal radius, even though the predicted
profiles always tend to be slightly flatter than the data.

The situation can be improved significantly by adopting a 
weakly tangentially anisotropic velocity distribution for the stars, as shown in 
Figure~\ref{fig7}. Indeed, for $\beta=-1$ ({\it dashed lines}),
subhalos with $V_{\rm max}=25\kms$ are nicely
consistent with most of the data points for both Draco and Fornax. 
Note that a mildly tangentially anisotropic distribution of stellar orbits 
was required by \citet{kleyna_etal02} in their best fit models to the 
kinematics of Draco. Moreover, \citet{lokas02} finds that the steeper 
the dark halo inner density cusp, the more tangential 
a stellar velocity distribution is required to fit the data
for both Draco and Fornax. Interestingly, modeling based on high-resolution 
Keck spectroscopy of six Virgo dwarf elliptical galaxies having an 
average line-of-sight velocity dispersion in the range $\sigma \sim 25-50\kms$ 
also yields for most of them best fit models to their 
kinematics that are consistent with mildly tangentially anisotropic orbits
\citep{geha_etal02}. Radially anisotropic models with $\beta=0.5$ ({\it dotted lines}) 
overestimate the central stellar velocity dispersion by a factor of 
$\sim 2$ and lead to steeply declining velocity dispersion profiles,
contrary to the trend in the observational data.
As in the case of the isotropic orbits, if we assume  
a tidal radius $3$ times larger ({\it right}) the match to 
the stellar velocity profiles becomes better.

\begin{figure*}
\centerline{\epsfxsize=3.8in \epsffile{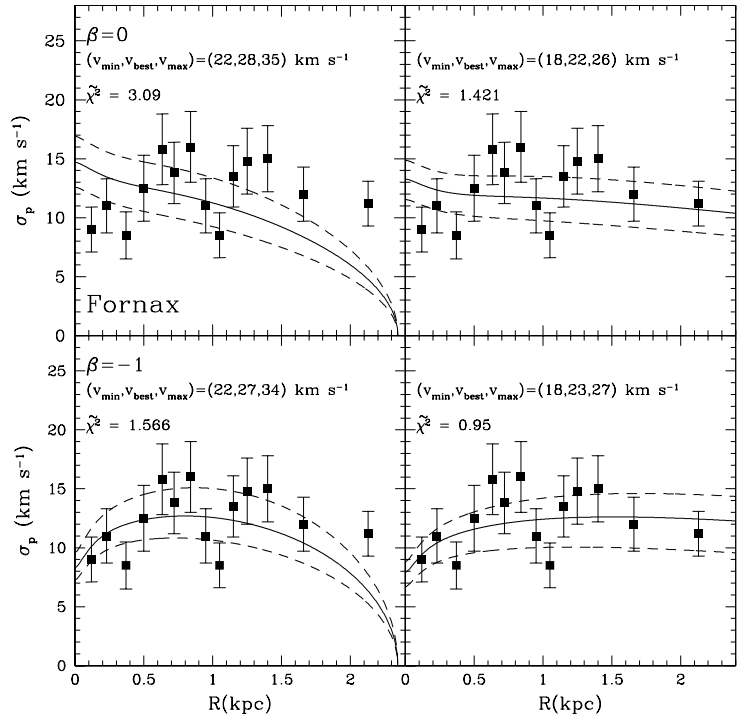}}
\caption{Same as Figure~\ref{fig8}, but for the dSph Fornax.
Similarly to the case of Draco, subhalos with $V_{\rm max}$ in the range 
$20-30 \kms$ provide the best fit to the data (solid lines), 
while $V_{\rm max} > 35 \kms$ are  $3\sigma$ (dashed lines) or more away from 
the best fit.
\label{fig9}}
\end{figure*}

In Figures~\ref{fig8} and~\ref{fig9}, we present the best 
fit models ({\it solid lines}), together with the $3\sigma$ intervals 
({\it dashed lines}) for Draco and Fornax respectively, assuming isotropic 
({\it top}) and tangentially anisotropic ({\it bottom}) models. In each case, we fix the 
value of the concentration parameter and fit, by $\chi^2$ minimization the maximum circular
velocity $V_{\rm max}$ with the two different values for the
tidal radius of the stellar component. Note that for Draco the best fit is calculated taking into
account that the error bars are asymmetric. The fixed concentration parameter
we used ($c=21$) is consistent with nearly all the profiles of the subhalos in
S02. Higher concentration models would imply lower values of $V_{\rm max}$, and
therefore our choice is conservative. This more detailed 
analysis shows that subhalos with $V_{\rm max}$ in the
range $20-35 \kms$ provide the best fit to the data 
and also that the high values favoured by S02, $V_{\rm max} > 40 \kms$,
are  $3\sigma$ or more away from the best fit. Determining the absolute
best fit to the data requires fitting more parameters, which is clearly
outside the scope of the present paper \citep[however, see][]{lokas02}.
In general, the degree of anisotropy can be
a function of radius instead of being simply a constant parameter. In
models in which the dSphs are formed by the tidal stirring and
transformation of disk-like systems similar to present-day dwarf irregular
galaxies \citep{mayer_etal01a,mayer_etal01b}, the remnant is a triaxial system 
whose velocity anisotropy depends on the radius. The remnants
produced in those simulations can be fitted by either an exponential or 
a King profile, similar to the observed dSphs.
Although there is considerable scatter in their structural properties 
\citep{mayer_etal01b}, in general the stellar orbits are tangentially anisotropic 
in the central regions (at distances equivalent to the core radii of 
Draco and Fornax) and become nearly isotropic or slightly radially anisotropic in the 
outer regions, where they are partially flattened by rotation. The interesting point is
that values of $\beta$ ranging from $-0.5$ to $-1.3$ are typical in the region
where most of their bound mass is. This result is confirmed by
recent gas-dynamical simulations with radiative cooling and heating and star
formation \citep[L. Mayer \etal 2004, in preparation;][]{mayer_wadsley03}
and lends support to the results shown in Figure~\ref{fig7} for $\beta = -1$.
Note that \citet{zentner_bullock03} suggested that a mild radial anisotropy 
for the dSphs equal to $\beta =0.15$ should be expected because the latter value
is typical of the central regions of simulated CDM halos, where the dwarfs 
actually reside. However, our results highlight that this is not necessarily the 
case. The tidally stirred stellar components develop a tangential 
anisotropy even though they evolve inside an initially isotropic 
dark matter halo because their evolution is governed by non-axisymmetric,
bar/buckling instabilities \citep{mayer_etal01b}.

We end by emphasizing that if the density profiles of the subhalos
models are initially steeper than an NFW profile by either having 
intrinsically steeper inner slopes (as might be possible for some
of the satellites in our cosmological runs; see \S~\ref{subsection:cosmo_sim}) 
or becoming steeper as a result of the adiabatic contraction following the 
collapse of baryons \citep{blumenthal_etal86}, circular velocities even lower
than the above values would be required to acceptably reproduce the observed 
data.

\section{DISCUSSION}
\label{section:discussion}

The results presented in the previous sections indicate that the conversion between 
$\sigma_{\star}$ and $V_{\rm max}$ initially adopted by \citet{moore_etal99} and 
\citet{klypin_etal99} is reasonable. Low values of  
$V_{\rm max}$ are required if the dSphs are embedded in CDM halos.
There is a second, phenomenological reason why the observed satellites should
not reside in halos as massive as the ones postulated by S02.
In this case, their baryonic components would lie
within a very small region corresponding to only $1\%$ of the 
virial radius of the object, posing a problem for structure formation 
scenarios. Indeed, in the prevailing galaxy formation paradigm 
\citep[e.g.,][]{fall_efstathiou80,mo_etal98}, the baryons condense into a 
rotationally supported structure whose size is determined by the initial spin 
parameter $\lambda$ and concentration $c$ of the dark matter halo, and
the fraction of mass and angular momentum of baryons relative to 
that of the halo. Assuming that the specific angular momentum of baryons is conserved 
during their infall and that the halo and baryons start with the same specific
angular momentum \citep{mo_etal98}, one can easily show that for a halo as massive as 
$V_{\rm vir}=50\kms$, a spin parameter equal to $\lambda <0.01$ would have been needed 
for the baryons to collapse to the observed size of Draco.
However, such a small value for the spin occurs in less than $1\%$ of the 
cosmological halos \citep{warren_etal92,lemson_kauffmann99,gardner01}.

The distribution of spin parameters has been measured so far only for isolated
halos and at mass scales larger than those of dwarf galaxies 
\citep[e.g.,][]{bullock_etal01b,gardner01}.
Although a new systematic analysis overcoming the above
limitations will be needed in the future, we made the first step forward
in this direction and we measured the spin parameter $\lambda$ for some of our 
subhalos. In particular, we selected ten isolated halos from the high resolution
region of run R6 and six subhalos within the virial radius of the cluster
in both runs R6 and R12. All halos are selected by mass 
($M_{\rm vir} \simeq 6.3 \times10^{10} \,h^{-1} \Mo$). The spin
parameters are measured at a radius equal to $0.5 \,r_{\rm vir}$
for the isolated halos, where $r_{\rm vir}$ denotes the halo virial radius,
and at $0.7 \,r_{\rm t}$ for the subhalos, where $r_{\rm t}$ denotes 
the subhalo tidal radius. The $\lambda$-values we found for both isolated halos and 
subhalos are between $0.024$ and $0.12$ and follow a log-normal distribution 
with $\overline{\lambda} = 0.039$ and $\sigma_{\lambda} = 0.55$
in agreement with previous studies using a much larger halo sample
\citep[e.g.,][]{bullock_etal01b}. This finding confirms our phenomenological 
argument that the dSph satellites could not be embedded in very massive halos.
Finally, it is interesting to note that the comparison of the 
spin parameters of the same subhalos in runs R6 and R12 shows that
they are independent of resolution.
Removal of baryons from an initially more extended system by tidal 
stripping, by supernova feedback, or even by photoevaporation from 
the UV background cannot be invoked to reduce the size of 
the baryonic component of the dwarfs at a later stage, because the halo 
would have been massive enough to suppress all these mechanisms 
\citep{benson_etal02b}. One then would have to rely on some catastrophic loss of 
angular momentum during baryonic collapse, but there is no obvious 
reason why this should have happened. 

Recently, \citet{kravtsov_etal04} proposed a qualitatively different 
solution to the missing satellites problem. Using high
resolution cosmological simulations of the formation of a Milky Way
sized halo in a {\LCDM} universe, these authors argued that 
the luminous dSphs in the Local Group can be identified with 
halos that had considerably higher masses and circular velocities when they 
formed at high redshift. Their model provide a convincing explanation
as to why the dSphs could retain their gas and form stars after
reionization and indicates that they are embedded in dark matter halos
that suffered dramatic mass loss due to tidal stripping.
We believe that the major difference between the subhalos' density 
structure found in our study and that claimed by S02 is due to resolution effects
affecting the determination of the circular velocity.
On the other hand, the reduced concentration that H03 find for their simulated 
satellites may be an artifact of non-equilibrium initial conditions combined with
numerical resolution, rather than reflecting the influence of tidal shocks
on the inner regions of the satellites.
The evolution of the internal structure is expected to be 
dramatically different for models that are not self-consistent compared to 
models for which one considers carefully the exact dynamics. 
Regarding this issue, \citet{kazantzidis_etal04} found that NFW satellites 
initialized using the Maxwellian approximation experience artificially accelerated 
mass loss and can completely disrupt in an external tidal field. The same satellites,
however, survive in identical experiments once the exact self-consistent velocity 
distribution of the model is taken into account.

Of course, one cannot exclude that some dynamical mechanism other than 
tidal interactions might have been responsible for lowering the concentration of the
subhalos. For example, if bars are effective in redistributing the angular momentum 
between the baryons and the halo \citep{weinberg_katz02}, then the dSphs could have been 
strongly affected by this mechanism if they originate from the morphological evolution 
of tidally stirred dwarf irregular galaxies. Indeed, tidal stirring always incorporates
a strong bar instability phase. However, recent work 
suggests that the amount of angular momentum that a bar transfers to the halo is 
not enough to turn the inner cusp into a core \citep{sellwood03}.

The results presented in this paper argue that there is no simple solution to the 
substructure problem within CDM models. The kinematics of the dSphs
are indeed a reasonable tracer of their dark halo potential wells.
However, their kinematics can also be accurately reproduced with 
dark halo density profiles that are nearly flat in the center
(see Figures~\ref{fig5} and~\ref{fig6}).
The latter fact illustrates the need to use better quality data and perhaps
higher order moments of the velocity distribution to break this degeneracy
and constrain the central cusp slopes \citep{lokas02}.
Our findings also have important implications for indirect dark matter
detection experiments attempting to observe the resulting
$\gamma$-ray flux from neutralino annihilation.
The expected flux depends sensitively
on the line-of-sight integral of the square of the 
mass density. This integral changes by more than one order of magnitude 
between the NFW and the much shallower density profiles found by S02,
as pointed out by the same authors in a more recent paper \citep{stoehr_etal03}.
With such shallow density profiles, Air-Shower-Cerenkov telescopes 
like {\sc Veritas} and {\sc Magic} might barely detect the emission from the 
Galactic center and would completely miss that coming from the dSphs. 
However, for dark matter particle candidates with properties consistent 
with some of the minimal supersymmetric models \citep{bergstrom_etal98} and 
with the persistently steep inner density profiles found in this paper, the Galactic 
center would be easily detected and the dSph satellites still
represent potentially valuable targets \citep{calcaneo_moore00}.

\section{SUMMARY}
\label{section:summary}

We have investigated the structural evolution of substructure,
using a set of high-resolution cosmological $N$-body 
simulations, coupled with simulations of the tidal stripping of individual satellites 
orbiting within a static host potential and employing up to $N=10^{7}$ particles.
Our main results and conclusions may be summarized as follows:

\begin{enumerate}
\item Cuspy satellite halos on eccentric orbits do not experience significant 
mass redistribution in their centers, and they retain the same steep inner 
density slope even after several strong pericentric tidal shocks.
This result is valid for our high-resolution cosmological simulations, in which 
the satellites evolve within a time-dependent cosmological 
tidal field, and for our simulations of individual subhalos
orbiting within a massive host system modeled as a fixed potential.

\item Convergence in the circular velocity profiles occurs much more slowly 
than that of the density profiles. This is due to the fact that low central 
resolution propagates to larger radii on account of the cumulative nature of the
former quantity. This discourages any attempt to use circular velocities of 
simulated satellites to derive their structural properties.

\item The density structure of tidally stripped cuspy halos 
can be well approximated by a simple fitting function (eq. [5]),
comprising an unmodified power-law central slope and an exponential cutoff, 
that describes the satellite's boundary imposed by the tides.

\item The predicted kinematics of the dSph galaxies Fornax and Draco,
assuming that they are embedded in the potential wells of our tidally stripped 
satellites, were compared to the observed stellar velocity dispersion profiles. 
Adopting isotropic and tangentially anisotropic models for the 
stellar component, we find that dark matter halos with maximum circular 
velocities in the range $V_{\rm max} \sim 20-35 \kms$ fit the data better, 
whereas models with $V_{\rm max} \gtsim~40\kms$ are at least at the 
$3\sigma$ level from the best fit values. If the initial central density
slopes of the simulated subhalos were steeper than the ones adopted here, even lower values
of $V_{\rm max}$ would be required to fit the data.
Tidal interactions do not provide the mechanism
for embedding the luminous dwarf galaxies of the Milky Way 
within the most massive subhalos in a {\LCDM} universe,
and, therefore, for reconciling the overabundance of Galactic 
satellites with CDM predictions.

\item Models with a mild tangential anisotropy of stellar orbits reproduce better
the shape of the observed velocity dispersion profiles of both Fornax and Draco,
whereas radially anisotropic models overestimate the central stellar velocity 
dispersion by a factor of $\sim 2$. Such a tangential anisotropy is expected in scenarios in 
which the dSphs result from the tidal stirring of systems similar to present-day 
dwarf irregular galaxies.
\end{enumerate}
\acknowledgments

We would like to thank the referee, Fabio Governato, for constructive comments on the
manuscript and Francisco Prada for organizing the La Palma Cosmology Conference 
``Satellites and Tidal Streams'' (2003) which motivated some of this work. 
Stimulating discussions with Andrey Kravtsov and Simon White are acknowledged. 
We are also grateful to Rocco Piffaretti for assistance with the statistical analysis 
and to the Swiss Center for Scientific Computing (CSCS), where the generation of the 
initial conditions for the cosmological runs was performed. 
The numerical simulations were carried out on the 
zBox supercomputer\footnote{See http://www-theorie.physik.unizh.ch/$\sim$stadel/}
and on LeMieux at the Pittsburgh Supercomputing Center.

\end{document}